\documentclass[graybox]{svmult}
\usepackage{mathptmx}       
\usepackage{helvet}         
\usepackage{courier}        
\usepackage{type1cm}        
\usepackage{makeidx}         
\usepackage{graphicx}        
\usepackage{multicol}        
\usepackage[bottom]{footmisc}

\makeindex             

\begin{document}

\title*{New Neighbours: Modelling the Growing Population of $\gamma$-ray Millisecond Pulsars}
\titlerunning{Modelling $\gamma$-ray Millisecond Pulsars}
\author{C. Venter, A.K. Harding, and T.J. Johnson}
\institute{C. Venter \at Unit for Space Physics, North-West University, Potchefstroom Campus, Private Bag X6001, Potchefstroom 2520, South Africa
\and
A.K. Harding and T.J. Johnson \at Astrophysics Science Division, NASA Goddard Space Flight Center, Greenbelt, MD 20771, USA}
%
%
\maketitle

\abstract*{The \textit{Fermi} Large Area Telescope, in collaboration with several groups from the radio community, have had marvellous success at uncovering new $\gamma$-ray millisecond pulsars (MSPs). In fact, MSPs now make up a sizable fraction of the total number of known $\gamma$-ray pulsars. The MSP population is characterized by a variety of pulse profile shapes, peak separations, and radio-to-$\gamma$ phase lags, with some members exhibiting nearly phase-aligned radio and $\gamma$-ray light curves (LCs). The MSPs' short spin periods underline the importance of including special relativistic effects in LC calculations, even for emission originating from near the stellar surface. We present results on modelling and classification of MSP LCs using standard pulsar model geometries.} 

\abstract{The \textit{Fermi} Large Area Telescope, in collaboration with several groups from the radio community, have had marvellous success at uncovering new $\gamma$-ray millisecond pulsars (MSPs). In fact, MSPs now make up a sizable fraction of the total number of known $\gamma$-ray pulsars. The MSP population is characterized by a variety of pulse profile shapes, peak separations, and radio-to-$\gamma$ phase lags, with some members exhibiting nearly phase-aligned radio and $\gamma$-ray light curves (LCs). The MSPs' short spin periods underline the importance of including special relativistic effects in LC calculations, even for emission originating from near the stellar surface. We present results on modelling and classification of MSP LCs using standard pulsar model geometries.}

\section{Introduction}
\label{sec:Intro}
To date, \textit{Fermi} Large Area Telescope (LAT) has uncovered more than a dozen new $\gamma$-ray millisecond pulsars (MSPs)~\cite{Harding10}. These MSPs were discovered by folding the $\gamma$-ray data with the known radio period, using timing solutions provided by various radio telescopes.
Discovery of the first eight \textit{Fermi}-LAT MSPs~\cite{Abdo09_MSP} indicated that these MSP pulse profiles and spectral properties mimic those of the younger pulsars, and that there are non-zero lags between the radio and $\gamma$-ray peaks. However, the discovery of PSR~J0034$-$0534~\cite{Abdo09_J0034} exhibiting phase-aligned $\gamma$-ray and radio peaks provided evidence for co-location of the radio and $\gamma$-ray emission regions. This phenomena has previously only been observed for the Crab pulsar. Recent detection of $\gamma$-ray pulsations aligned with the radio peaks~\cite{Ray10,Abdo10_2msps} from the famous `first' MSP PSR~J1939+2134~\cite{Backer82}, as well as the first `black-widow' MSP PSR~J1959+2048~\cite{Fruchter88}, now provide two more examples of this exceptional behaviour. Publication of a number of new $\gamma$-ray MSP detections are awaited~\cite{Ransom10} (some resulting from follow-up radio observations of unpulsed \textit{Fermi}-LAT sources which appear in the First \textit{Fermi}-LAT source catalogue), and the growing MSP population now constitute a significant fraction of all \textit{Fermi}-LAT pulsar detections to date~\cite{Abdo09_Cat}.

The $\gamma$-ray MSPs exhibit a variety of light curve (LC) shapes, and have been modelled using standard two-pole caustic (TPC), outer gap (OG), and pair-starved polar cap (PSPC) models~\cite{Venter_MSP09}, an annular gap model~\cite{Du10}, lately also altitude-limited TPC / OG models~\cite{Venter10}, and lastly using a force-free magnetic field~\cite{Bai09}. We elaborate on some of these models in Section~\ref{sec:Models}. (See~\cite{Harding10} for a review of $\gamma$-ray pulsar population modelling.)


\section{Pulsar Models}
\label{sec:Models}

Below is a brief description of some modelling aspects (see~\cite{Venter_MSP09} for details).
\subruninhead{B-field} As before, we assume the retarded vacuum dipole field~\cite{Deutsch55,Dyks04_Bovc} to be the basic geometric structure of the MSP magnetosphere. The magnetic moment is inclined by an angle $\alpha$ with respect to the spin axis, while $\zeta$ is the observer angle.
\subruninhead{Caustics} Due to the rapid increase of the co-rotation velocity with altitude, approaching $c$ as one reaches the light cylinder, it is important to include relativistic effects when calculating LCs. The effects of (i) B-field line curvature, (ii) time-of-flight delays, and (iii) aberration combine to form intense caustic emission on the trailing field lines, while the emission is spread out on the leading field lines~\cite{Morini83,Dyks04_B}.
\subruninhead{Polar Cap (PC) Model} The PC model~\cite{Daugherty82} assumes extraction of primary charges from the stellar surface, acceleration of these charges to relativistic velocities along B-field lines, and subsequent $\gamma$-ray curvature radiation and pair creation at a very low altitude. A pair-formation front (PFF) is formed which screens the accelerating E-field ($E_{||}$) parallel to the B-field.
\subruninhead{Faded PC Model} This is a geometric approximation~\cite{Venter10}, where we use a rising and falling exponential function to modulate the assumed emissivity close to the stellar surface, as motivated by pair cascade modelling.
\begin{figure}[t]
\includegraphics[scale=.63]{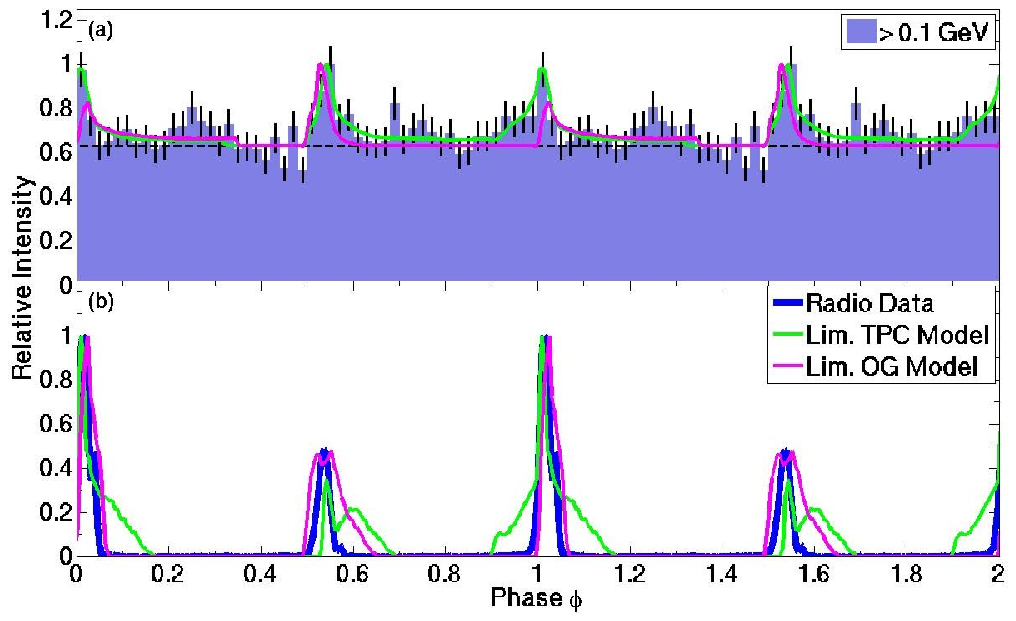}
\vskip0.2cm
\includegraphics[scale=.63]{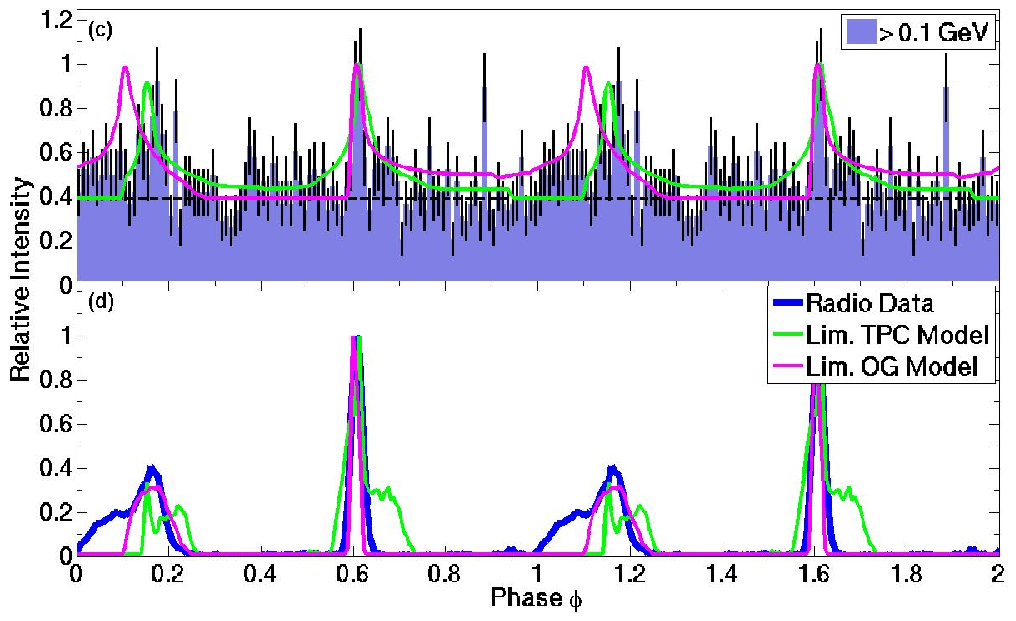}
\caption{Altitude-limited TPC / OG LC fits for PSR~J1939+2134 (panel~[a]: \textit{$\gamma$-rays}; panel~[b]: \textit{radio}) with $(\alpha^{\rm TPC},\zeta^{\rm TPC})=(65^\circ,80^\circ)$, $(\alpha^{\rm OG},\zeta^{\rm OG})=(70^\circ,90^\circ)$, $(R^{\rm \gamma,TPC}_{\rm min},R^{\rm \gamma,TPC}_{\rm max})=(0.12,1.05)R_{\rm LC}$, $(R^{\rm \gamma,OG}_{\rm min},R^{\rm \gamma,OG}_{\rm max})=(0.12,1.00)R_{\rm LC}$, $(R^{\rm radio,TPC}_{\rm min},R^{\rm radio,TPC}_{\rm max})=(0.93,1.20)R_{\rm LC}$, $(R^{\rm radio,OG}_{\rm min},R^{\rm radio,OG}_{\rm max})=(0.75,1.10)R_{\rm LC}$. Fits for PSR~J1959+2048 (panel~[c]: \textit{$\gamma$-rays}; panel~[d]: \textit{radio}) with $(\alpha^{\rm TPC},\zeta^{\rm TPC})=(60^\circ,80^\circ)$, $(\alpha^{\rm OG},\zeta^{\rm OG})=(55^\circ,85^\circ)$, $(R^{\rm \gamma,TPC}_{\rm min},R^{\rm \gamma,TPC}_{\rm max})=(R^{\rm \gamma,OG}_{\rm min},R^{\rm \gamma,OG}_{\rm max})=(0.12,1.20)R_{\rm LC}$, $(R^{\rm radio,TPC}_{\rm min},R^{\rm radio,TPC}_{\rm max})=(0.95,1.05)R_{\rm LC}$, $(R^{\rm radio,OG}_{\rm min},R^{\rm radio,OG}_{\rm max})=(0.90,1.10)R_{\rm LC}$.}
\label{fig:2}       
\end{figure}
\subruninhead{PSPC Model} When $E_{||}$ is not large enough to accelerate primary electrons to sufficiently high energies so pair cascades will occur, it will be unscreened, and one expects acceleration and radiation from the full open field line region above the PC~\cite{HM98,HUM05}.
\subruninhead{SG / TPC Model} When enforcing the boundary condition that $E_{||}$ vanishes on the last open B-field lines (LOFLs), the PFF altitude becomes dependent on magnetic co-latitude, and narrow gaps form just inside of the LOFLs~\cite{Arons83}. The SG model~\cite{MH03_SG} may provide the framework for the geometric TPC model \cite{Dyks03}.
\subruninhead{OG Model} The Goldreich-Julian charge density~\cite{GJ69} changes sign at the null charge surface (NCS), and OGs are assumed to form close to the LOFLs, above the NCS~\cite{CHR86a,Romani96}.
\subruninhead{Altitude-limited TPC / OG Models} We lastly investigated geometric TPC / OG models with arbitrary lower, $R_{\rm min}$, and upper, $R_{\rm max}$, radial gap extensions~\cite{Abdo09_J0034}.

\section{Results}
\label{sec:Results}
\subruninhead{First 8 MSPs} We found~\cite{Venter_MSP09} that~6 of the~8 \textit{Fermi}-LAT MSPs may be modelled using standard TPC / OG models, while~2 are exclusively fit by PSPC models.
\subruninhead{PSR~J0034$-$0534} We found altitude-limited TPC / OG LC fits~\cite{Abdo09_J0034}, with $(\alpha,\zeta)=(30^\circ,70^\circ)$, $(R^\gamma_{\rm min},R^\gamma_{\rm max})=(0.12,0.9)R_{\rm LC}$, and $(R^{\rm radio}_{\rm min},R^{\rm radio}_{\rm max})=(0.6,0.8)R_{\rm LC}$, with $R_{\rm LC}=cP/2\pi$, and $P$ the pulsar period. There also exist solutions~\cite{Venter10} for the faded PC model for $(\alpha,\zeta)=(10^\circ,\sim34^\circ)$,
although the first ones seem more credible.
\subruninhead{PSR~J1939+2134 and PSR~J1959+2048} Figure~\ref{fig:2} indicates altitude-limited TPC / OG LC fits with aligned radio and $\gamma$-ray pulses~\cite{Abdo10_2msps}. 

\section{Discussion and Conclusions}
\label{sec:Con}
Our results indicate that the high-energy (HE) radiation comes from narrow gaps in the outer magnetosphere, requiring the presence of robust pair formation to screen much of the open field line region. However, MSPs do not produce enough pairs in the standard vacuum dipole models to screen $E_{||}$. Standard geometrical models still provide suitable LC fits though, and the resulting $\alpha$ and $\zeta$ are in reasonable agreement with fits from radio polarization measurements. In the case of phase-aligned $\gamma$-ray and radio pulses, we infer the existence of co-located emission regions, as well as high-altitude radio emission. Future work includes investigations of HE spectra using full radiation codes, collective emission from MSPs in globular clusters, and pulsar population studies.

\begin{acknowledgement}
CV is supported by the South African National Research Foundation. AKH acknowledges support from the NASA Astrophysics Theory Program.
\end{acknowledgement}

\end{document}